\documentclass[reprint,superscriptaddress,amsmath,amssymb,aps,prb]{revtex4-2}
\usepackage{amsmath, amsfonts}
\usepackage{float}
\usepackage{graphicx} 
\usepackage{dcolumn} 
\usepackage{bm} 
\usepackage{xcolor}
\usepackage{array}
\usepackage{tabularx}

\usepackage[colorlinks=true,linkcolor=blue,urlcolor=blue,citecolor=blue]{hyperref}

\begin{document}

\title{Microscopic Theory of Superionic Phase Transitions: \\ Nonadiabatic Dynamics and Many-Body Effects}


\author{Jiaming Hu}
\email{hujiaming@westlake.edu.cn}
\affiliation{Department of Materials Science and Metallurgy, University of Cambridge, Cambridge CB3 0FS, United Kingdom}
\affiliation{School of Engineering, Westlake University, Hangzhou 310030, China.}

\author{Zhichao Guo}
\affiliation{Center for Quantum Matter, Zhejiang University, Hangzhou 310058, China.}

\author{Jingyi Liang}
\affiliation{Department of Geography, The University of Hong Kong, Hong Kong, China.}

\author{Bartomeu Monserrat}
\email{bm418@cam.ac.uk}
\affiliation{Department of Materials Science and Metallurgy, University of Cambridge, Cambridge CB3 0FS, United Kingdom}


\date{\today}
\begin{abstract}
    Superionic phase transitions have attracted extensive interest for decades due to their promising applications and rich underlying physics. In particular, complicated many-body effects and nonadiabatic dynamics are believed to play essential roles, limiting the explanatory power of phenomenological approaches and obscuring the microscopic mechanisms at play. In this work, we develop a unified theoretical framework for describing solid-state ionic conduction. After reviewing the conventional approximations, we construct a general lattice model that applies to both normal ionic and superionic conductors. By incorporating the nonadiabatic concerted-hopping mechanism and the many-body Coulomb interaction within a self-consistent mean-field scheme, we identify these two effects as the fundamental driving forces behind type-I and type-II superionic phase transitions, respectively. Our model directly reproduces key experimental observations. Within this unified framework, we further provide a comprehensive comparison between the two types of transitions. Overall, our work offers microscopic insight into superionic phase transitions and provides guidance for the design and optimization of advanced solid-state ionic conductors. 
\end{abstract}

\maketitle


\section{Introduction}\label{sec:introduction}

Superionic conductors (SICs) are materials with exceptionally high ionic conductivity, which have diverse applications in solid-state electrolytes~\cite{review_applications, review_SSEs, goodenough2015solid, jay2015genetics, famprikis2019fundamentals,morgan2021understanding}, thermoelectric energy conversion~\cite{liu2012copper,snyder2004disordered,chalfin2007cation,ren2023extreme}, and neuromorphic technologies~\cite{PhysRevLett.130.268401,li2025situ}. Many SICs can be activated from a low-conductivity phase (normal phase) to a high-conductivity phase (superionic phase) when temperature rises beyond a material-dependent value~\cite{boyce1979superionic,hull2004superionics,article}. This superionic phase transition is generally categorized into two types: type-I is a first-order transition, characterized by an abrupt change in ionic conductivity, and is exemplified by silver and cuprous halides and chalcogenides~\cite{funke1976agi}, as well as their extensions like $\alpha$-KAg$_3$Se$_2$~\cite{rettie2021two}; type-II is a second-order transition, with a continuous but non-smooth change in ionic conductivity, and often follows distinct Arrhenius relationships in the normal and superionic regimes (e.g. $\alpha$-Li$_3$N~\cite{krenzer2023nature}, MCrX$_2$, M=Cu,Ag, X=S,Se~\cite{BOUKAMP19831193,pnas}).


In recent decades, extensive experimental characterizations and numerical simulations have been conducted to investigate SICs~\cite{wu2026ab, zhang2020abnormally,ren2023extreme,rettie2021two}. A variety of theoretical models have also been proposed, including the free-ion model~\cite{maierBook}, lattice-gas model~\cite{lattice_model}, continuous stochastic model~\cite{stochastic_model}, and percolation model~\cite{percolation_model}. Diverse mechanisms have also been suggested to explain superionic behavior, such as order–disorder transitions~\cite{zhang2019superionic,eapen2017entropic,wang2021atomistic,niedziela2019selective,hu2021observation,bernstein2012origin}, cation-sublattice partial melting~\cite{zheng2021antiperovskite}, competition between local atomic preferences (frustration)~\cite{wood2021paradigms}, overpopulation of lattice sites (jamming)~\cite{bernstein2012origin,annamareddy2017low}, and the paddle-wheel mechanism~\cite{zhang2022exploiting}. Moreover, significant changes in vibrational spectra have been observed across superionic phase transitions~\cite{niedziela2019selective,pnas,gupta2022strongly}, highlighting the crucial role of anharmonic vibrations in superionic conduction, as further supported by recent materials discovery studies~\cite{lopez2023universal,muy2018tuning,muy2019high,lopez2023universal,krenzer2022anharmonic}.

Based on these efforts, it has been found that type-I superionic phase transitions reflect a fundamental change in the transport mechanism, whereas type-II transitions are primarily associated with variations in formation energetics of the ionic charge carriers. However, a clear gap remains between theoretical descriptions and experimental observations: existing studies have provided mostly phenomenological and case-dependent interpretations, lacking a unified explanation for the general mechanism and critical behavior near the transition point. In particular, the relationship between type-I and type-II superionic phase transitions remains ambiguous due to the absence of a unified theoretical framework. 

The main challenge in developing a theoretical framework arises due to the need to incorporate many-body effects and nonadiabatic dynamics. The former primarily originates from Coulomb interactions among mobile ions, which can actively drive collective ionic modes~\cite{MD_SIC,book_SIC}. The latter arises from the comparable dynamics between mobile species and host lattice in certain SICs~\cite{MD_SIC,niedziela2019selective,krenzer2022anharmonic}, which compromises the validity of the conventional adiabatic treatment of the host-lattice potential landscape. These limitations continue to impede a complete microscopic understanding of the mechanisms underlying superionic phase transitions.

In this work, we present a microscopic theory of superionic phase transitions. We first review in Sec.\,\ref{sec:approximations} the approximations commonly employed for conventional solid-state ionic conductors and assess their validity in the context of SICs. Building upon broadly applicable approximations, we then develop in Sec.\,\ref{sec:statistical_description} a statistical description of general inter-site hopping. Based on this framework, a unified lattice model applicable to both normal ionic conduction and SICs is constructed in Sec.\,\ref{sec:lattice_model_general}, with its reduction to the normal limit demonstrated in Sec.\,\ref{sec:lattice_model_conventional}. The reduced theory is subsequently applied to a one-dimensional system as a benchmark for normal ionic conductors in Secs.~\ref{sec:intrinsic_1D} and~\ref{sec:couple_E_1D}. In Secs.~\ref{sec:depol_meanfield_1D} and~\ref{sec:concert_meanfield_1D}, we go beyond the adiabatic and single-particle approximations by incorporating concerted-hopping effects and Coulomb interactions among mobile ions within a self-consistent mean-field framework, which naturally give rise to type-I and type-II superionic phase transitions, respectively.  
Finally, in Sec.\,\ref{sec:relation_meanfields_1D}, we show that the key results remain valid in higher-dimensional systems and provide a systematic analysis and summary of the relationships and distinctions between the two types of superionic phase transitions.


\section{Dynamics of mobile ions}

\subsection{Approximations and their validity}\label{sec:approximations}

\begin{figure}[htp]
    \centering
    \includegraphics[width=0.48\textwidth]{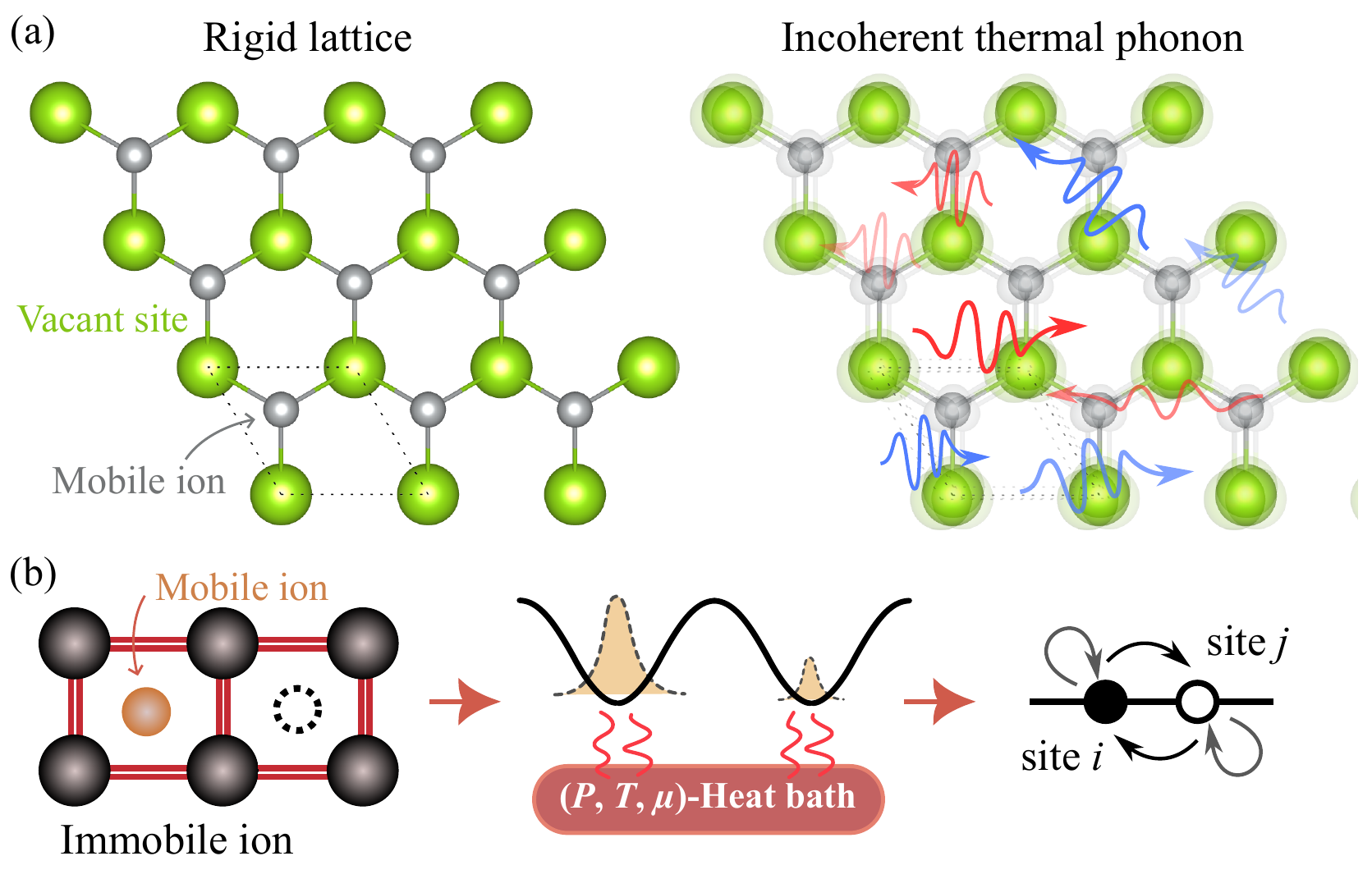}
    \caption{\label{fig:schematic_renormal}
    \textbf{Schematic of physical pictures for normal ionic conduction.} (a) Thermally excited crystal lattice as an incoherent superposition of thermal phonons. (b) Mobile ions that are confined within the adiabatic potential given by the host lattice do stochastic motion driven by the thermal excitation (schematicized as a heat bath). Tight-binding approximation further simplifies the potential landscape into lattice of sites and inter-site transitions. }
\end{figure}

We begin by reviewing the key approximations used to describe normal solid-state ionic conductors: 
\begin{enumerate}
\item \textit{Stochastic approximation}. At or near thermodynamic equilibrium, and neglecting quantum effects, the crystal lattice fluctuates around its equilibrium configuration due to the presence of thermal energy. This energy can be viewed as being carried by various mechanical waves that propagate and decay throughout the lattice with arbitrary relative phases, which is schematized in Fig.\,\ref{fig:schematic_renormal}(a). The superposition of these waves gives rise to the thermal motion of individual ions, which appears stochastic owing to the random phase relations among the waves. 

\item \textit{Adiabatic approximation}. The host sublattice formed by immobile ions typically undergoes sub-angstrom vibrations with characteristic frequencies in the terahertz regime corresponding to picosecond timescales. If these vibrations are much faster than the average waiting time between large-amplitude mobile-ion hops (typically 2-3~\AA)~\cite{poletayev2024persistence}, this separation of timescales justifies an adiabatic treatment in which mobile ions experience an effective potential generated by the averaged configuration of the host lattice. 

\item \textit{Tight-binding approximation}. The potential landscape established by the immobile ions defines a lattice of local potential minima for mobile ions, corresponding to the energetically stable \textit{sites}. Occupation of these sites represents long-lived states, while inter-site hops occur as transient events. Hence, as illustrated in Fig.\,\ref{fig:schematic_renormal}(b), it is possible to adopt a coarse-grained phase space where only site occupations are retained, and inter-site motion is represented as transitions between these site-occupation states. It is analogous to the tight-binding representation of electronic systems~\cite{book_advanced_solid_state_physics}. 

\item \textit{Single-particle approximation}. The mobile ions are often treated as independent from each other, an assumption valid only under conditions of strong screening or in the dilute limit. 
\end{enumerate}

Among these approximations, point (1) is generally valid as a fundamental principle of equilibrium (or near-equilibrium) thermodynamics. Point (3) is also generally valid, except for the type-I superionic phase where the mobile-ion sublattice becomes liquid-like, which will be addressed in Sec.\,\ref{sec:concert_meanfield_1D}. However, the other two approximations are not generally appropriate for SICs: point (2) may break down in concerted-hopping regimes~\cite{MD_SIC}, where inter-site hopping can be as active as the local vibrations of the host lattice; point (4) becomes invalid for strongly correlated or highly concentrated mobile-ion systems, where many-body effects are significant. 

Therefore, in the following three sections, we discuss the general statistical description of mobile-ion hopping; we build a general lattice model based only on the acceptable approximations in points (1) and (3); and we then simplify this model by adopting approximations (2) and (4) to reach the limit of normal ionic conductors. After that in Sec.\,\ref{sec:origins_of_superionic_phase_transition}, we go beyond approximations (2) and (4) to study superionic behaviors.


\subsection{Statistical description of mobile-ion hopping}\label{sec:statistical_description}

Consider a series of tight-binding sites for mobile ions, labeled by $j = 1, 2, \ldots, N_s$. The instantaneous occupation number at each site can only take binary values (0 or 1), while its ensemble average becomes a continuous variable which we refer to as the \emph{ionic density}. Specifically, the ionic density at time $t$ and site $j$, denoted by $n_j(t)$, represents a statistical average of the occupation number at site $j$ over all the possible evolution trajectories of the system, that is, the ensemble average. It is neither the quantum-mechanical probability in the sense of Born's statistical interpretation nor the real-space concentration in the continuum limit. 


Mobile-ion hopping is driven by thermal energy exchange with surrounding ions. Microscopically, mobile ions gain or release energy through interatomic forces as they approach or move away from neighboring ions while undertaking thermal vibrations, typically described in terms of phonons. The different vibrational modes, associated with vibrational frequency $\omega^{s,\bm{q}}$ and labeled by branch index $s$ and momentum $\bm{q}$, provide distinct channels for this energy exchange. The high temperatures typically present in SICs imply that the ionic vibrations are highly anharmonic, and $(s,\bm{q})$ should be considered as self-consistent phonon modes subject to a specific temperature~\cite{hooton1955li}. 

Each channel $(s,\bm{q})$ is characterized by its own temporal period $\tau^{s,\bm{q}}=1/\omega^{s,\bm{q}}$, periodically exchanging energy with mobile ions every time unit $\tau^{s,\bm{q}}$. Therefore, if an energy transfer through channel $(s,\bm{q})$ happens at moment $t$ for a given ionic density $\{n_j(t)\}$, it evolves as: 
\begin{equation}\label{eq:n_jsq_ttau}
    n_j^{s,\bm{q}}(t + \tau^{s,\bm{q}}) = {\sum\limits_i}P_{ji}^{s,\bm{q}}(t)n_i(t).
\end{equation}
Here $P_{ji}^{s,\bm{q}}(t)$ is the corresponding hopping probability from site $i$ to site $j$, which is given by Boltzmann statistics at thermodynamic equilibrium with temperature $T$ as: 
\begin{equation}\label{eq:P_ji_DOS}
    P_{ji}^{s,\bm{q}}(t) = \frac{1}{\lambda[E^{s,\bm{q}}_{ji}(t)]}\frac{{\int_{E^{s,\bm{q}}_{ji}(t)}^{+\infty}}e^{-E/k_{\rm B}T}g^{s,\bm{q}}(E;t)dE}{{\int_{E_{ii}}^{+\infty}}e^{-E/k_{\rm B}T}g^{s,\bm{q}}(E;t)dE}, 
\end{equation}
where $E^{s,\bm{q}}_{ji}$ is the minimum energy required for hopping from site $i$ to site $j$, $\lambda(E_{ji})$ is the degeneracy of available hopping destinations with the same threshold $E_{ji}$, and $E_{ii}$ is the chemical potential at the origin site $i$. The density of states $g^{s,\bm{q}}(E;t)$ incorporates the information about all the ionic degrees of freedom in the system. The time dependence in $E^{s,\bm{q}}_{ji}(t)$ and $g^{s,\bm{q}}(E;t)$ arises from the influence of the mobile ion density (e.g. many-body interactions), from nonadiabatic dynamics of the host lattice, or from an external field. For the simplest case of a trivial density of states $g(E)\sim{\rm constant}$ we have: 
\begin{equation}\label{eq:P_ji_simple}
    P_{ji}^{s,\bm{q}}(t) \rightarrow \frac{{\int_{E^{s,\bm{q}}_{ji}}^{+\infty}}e^{-E/k_{\rm B}T}dE}{{\int_{E_{ii}}^{+\infty}}e^{-E/k_{\rm B}T}dE} = e^{-\Lambda^{s,\bm{q}}_{ji}/k_{\rm B}T}, 
\end{equation}
where the activation energy $\Lambda^{s,\bm{q}}_{ji}=E^{s,\bm{q}}_{ji}-E_{ii}$ is the minimum hopping energy $E_{ji}$ with respect to the ground state energy of site $i$.

We can re-write these expressions more compactly by using the hopping matrix $\mathbb{P}^{s,\bm{q}}$ and ionic density vector $n^{s,\bm{q}}$, defined as $[\mathbb{P}^{s,\bm{q}}(t)]_{ji}\equiv{P}^{s,\bm{q}}_{ji}(t)$ and $n^{s,\bm{q}}\equiv(n^{s,\bm{q}}_1,n^{s,\bm{q}}_2,\ldots,n^{s,\bm{q}}_{N_s})^{\bm{T}}$, which lead to Eq.\,\eqref{eq:n_jsq_ttau} reducing to $n^{s,\bm{q}}(t + \tau^{s,\bm{q}}) = \mathbb{P}^{s,\bm{q}}(t)n(t)$. At later times $t + \tau$, the temporal evolution of this channel is given by: 
\begin{equation}
    \begin{aligned}
        n^{s,\bm{q}}(t + \tau)
        &= 
        \left[ \prod_{l=0}^{N_{\tau}^{s,\bm{q}}-1} \mathbb{P}^{s,\bm{q}}(t^{s,\bm{q}}_l) \right] n(t),
    \end{aligned}
\end{equation}
where $t^{s,\bm{q}}_l = t + l\tau / N_{\tau}^{s,\bm{q}}$ for $l = 0, 1, \ldots, N_{\tau}^{s,\bm{q}} - 1$ with $N_{\tau}^{s,\bm{q}} = [\tau/\tau^{s,\bm{q}}]$. We ignore the trivial case $\tau<\tau^{s,\bm{q}}$ since no evolution will happen. 

\begin{figure}[htp]
    \centering
    \includegraphics[width=0.44\textwidth]{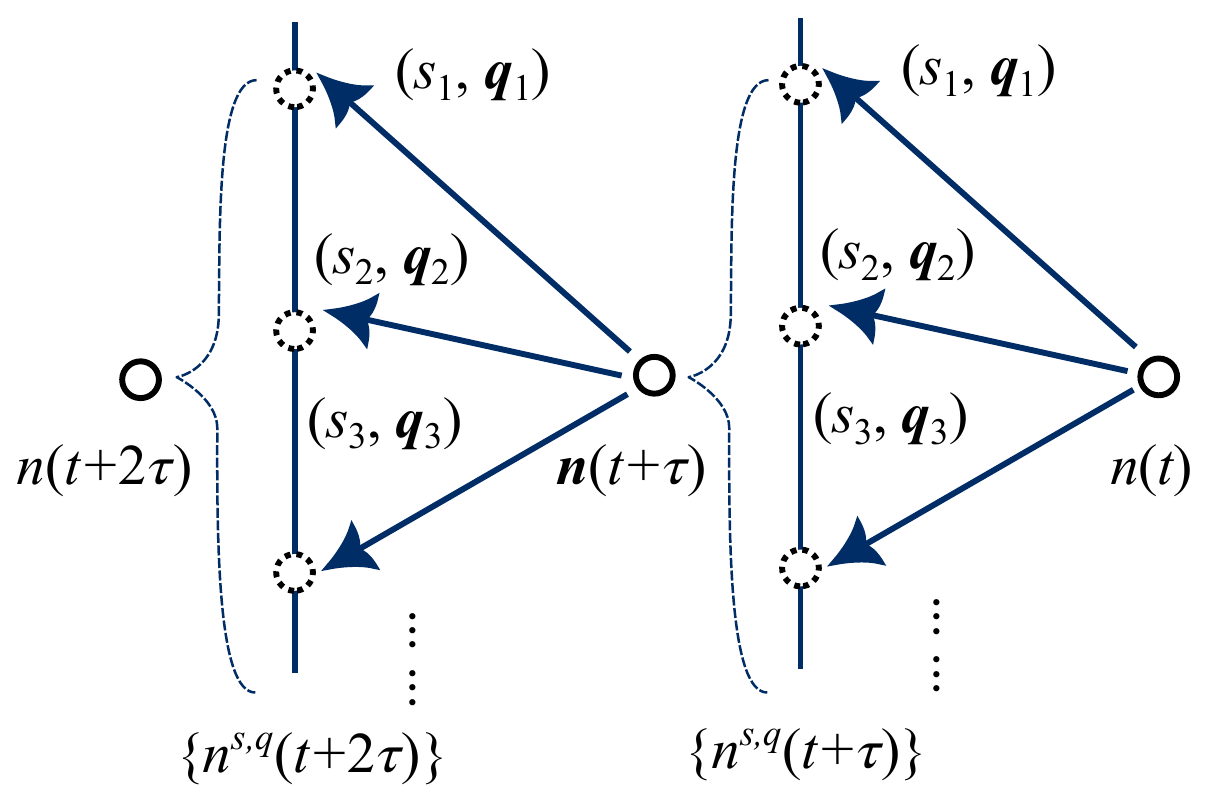}
    \caption{\label{fig:schematic_evolution}
    \textbf{Schematic of ensemble-averaged evolution along different phonon modes.} For each time step $t\rightarrow t+\tau$, the ionic density can evolve through thermal energy exchange with different phonon modes $(s_l,\bm{q}_l), l=1,2,\ldots$. The statistically observable evolution is then averaged over all possible channels. Note that the initial-phase configuration of Eq.\,\eqref{eq:trajectory_mode_initial_phase} is not explicitly schematized. }
\end{figure}

As schematized in Fig.\,\ref{fig:schematic_evolution}, the statistically observable evolution of the ionic density is the average over all possible ensemble realizations through all accessible channels, which is given by: 
\begin{equation}\label{eq:trajectory_mode_initial_phase}
    \begin{aligned}
        &n_{\{X^{s,\bm{q}}\}} (t + \tau)
        \\
        &= \frac{1}{N_{(s,\bm{q})}} \sum_{s,\bm{q}}^{N_{(s,\bm{q})}}
        \left[ \prod_{l=0}^{N_{\tau}^{s,\bm{q}}-1} \mathbb{P}^{s,\bm{q}}(t^{s,\bm{q}}_l + X^{s,\bm{q}})  \right] n(t),
    \end{aligned}
\end{equation}
where $N_{(s,\bm{q})}$ is the number of phonon modes. The time period $\tau$ should not be smaller than the minimum phonon vibrational period $\tau \ge 1/\max[\omega^{s,\bm{q}}]$. 
The offsets $X^{s,\bm{q}} \in [0, \tau^{s,\bm{q}})$ correspond to the relative initial phases, representing the fact that different channels may not deliver energy to the mobile ions synchronously. The notation $n_{\{X^{s,\bm{q}}\}}(t)$ emphasizes that the ionic density evolves under a specific initial phase configuration ${\{X^{s,\bm{q}}\}}$. 

If all modes are thermally populated, ${\{X^{s,\bm{q}}\}}$ can be regarded as independent random variables distributed as $X^{s,\bm{q}} \sim f(X^{s,\bm{q}})$, where $\int dx f(x) = 1$; for example one may set $f(X^{s,\bm{q}}) = 1 / \tau^{s,\bm{q}}$ under time translation invariance. This leads to the ensemble-averaged density: 
\begin{equation}
    n(t) =
    \left( \prod_{(s,\bm{q})}^{N_{(s,\bm{q})}} \int dX^{s,\bm{q}} f(X^{s,\bm{q}}) \right)
    n_{\{X^{s,\bm{q}}\}}(t).
\end{equation}
Therefore, the ionic density evolution can be formulated as $n(t + \tau) = \mathbb{P}(t)n(t)$ with the full hopping matrix $\mathbb{P}(t)$ defined as: 
\begin{equation}\label{eq:bbP_t_full}
    \begin{aligned}
        \mathbb{P}(t) &\equiv 
        \frac{1}{N_{(s,\bm{q})}} \sum_{s,\bm{q}}^{N_{(s,\bm{q})}} 
        \int dX^{s,\bm{q}} f(X^{s,\bm{q}})
        \\
        &{\times}
        \prod_{l=0}^{N_{\tau}^{s,\bm{q}}-1} \mathbb{P}^{s,\bm{q}}(t^{s,\bm{q}}_l + X^{s,\bm{q}})
    \end{aligned}
\end{equation}
The element-wise expression of this relation is simply given by: 
\begin{equation}\label{eq:njt_Pjit}
    n_j(t + \tau) = \sum_i^{N_s} P_{ji}(t)n_i(t).
\end{equation}
Note that ${\sum\limits}_{j}P_{ji}(t)=1$ since it accounts for all possible events, which in turn guarantees the conservation of the total mobile ion number as: 
\begin{equation}
    \begin{aligned}
        {\sum\limits_j^{N_s}}n_j(t + \tau) &= {\sum\limits_{i}^{N_s}}n_i(t){\sum\limits_j^{N_s}}P_{ji}(t) = {\sum\limits_i^{N_s}}n_i(t). 
    \end{aligned}
    \label{eq:number-conservation}
\end{equation}

The general expression for $P_{ji}(t)$ given in Eq.\,\eqref{eq:bbP_t_full} depends on the time $\tau$. Different choices of $\tau$ correspond to different effective dynamical patterns of inter-site hopping. A larger $\tau$ incorporates more low-frequency channels into the description, while simultaneously neglecting a greater amount of correlation between distinct channels. In the long-time limit $\tau \gg 1/\omega_m$, the influence of the initial phase becomes negligible, where $\omega_m$ denotes the lowest frequency among all modes that contribute to ionic hopping. In particular, if the inter-site hopping process of interest is dominated by a few phonon channels characterized by a typical frequency $\omega_c$, the hopping rate is expected to converge according to
\(
\sum_{j \neq i} P_{ji}(t)/\tau \rightarrow \text{constant} \quad \text{for} \quad \tau > 1/\omega_c,
\)
which represents an intrinsic property of the system. In the simplest case described by Eq.\,\eqref{eq:P_ji_simple}, it reduces to the widely adopted phenomenological form $\nu_0 e^{-E_{\rm A} / k_{\rm B} T}$, where $E_{\rm A}$ is the ``activation energy'' and $\nu_0$ denotes the ``attempt frequency''. This framework rationalizes the experimental observation that \emph{``ionic transport can be modeled as a random walk only at timescales longer than the persistence of correlations, but such a phenomenological random walk does not necessarily correspond to a true atomistic one''}~\cite{poletayev2024persistence}. Here we explicitly describe the transport as an ensemble-statistical process, suggests that macroscopic measurements cannot be directly interpreted in terms of atomistic quantities without properly accounting for this statistics. Our formulation provides a microscopic description derived from lattice dynamics.

Although, in principle, the summation in Eq.\,\eqref{eq:trajectory_mode_initial_phase} runs over all $N_{(s,\bm{q})}$ phonon modes, only modes with the following characteristics can make significant contributions:  

\begin{enumerate}
\item \textit{Appropriate eigenvector}. Only phonon modes whose eigenvectors exhibit substantial relative motion between mobile and immobile ions can efficiently mediate energy exchange between them, and thus play an important role in mobile-ion hopping. This feature manifests as a low value of $E^{s,\bm{q}}_{ji}(t)$. 

\item \textit{Moderate frequency}. As indicated by point~(1), the frequency of hopping-relevant modes reflects the strength of interatomic forces between the mobile ions and the host lattice. Consequently, a low frequency $\omega^{s,\bm{q}}$ typically corresponds to a small value of $E^{s,\bm{q}}_{ji}(t)$. On the other hand, a higher frequency implies a more rapid energy exchange, which facilitates more efficient thermal excitation. Therefore, as a trade-off, an intermediate value of $\omega_{s,\bm{q}}$ is generally preferred. This feature is supported by the observed correlation between phonon band centers and ionic hopping rates~\cite{krauskopf2018comparing}, and the relation between lowest optical frequency and activation energy~\cite{wakamura1997roles}. 

\item \textit{Weak dispersion}. Phonon modes with weak dispersion 
generally correspond to localized ionic vibrations with low group velocities. When such modes form wave packets induced by thermal fluctuations, they propagate slowly and tend to concentrate their energy on only a few ions before dissipating, thereby enhancing the probability of local ionic hopping. This feature is reflected in the density of states $g^{s,\bm{q}}(E;t)$.
\end{enumerate}
    
In practice, the extrema of optical phonon branches are often more likely to satisfy these conditions and are therefore expected to be the dominant driving force behind ionic motion. 



\subsection{Lattice model for general solid-state ionic conductors}\label{sec:lattice_model_general}

The mode-resolved structure of $P_{ji}(t)$ in Eq.\,\eqref{eq:bbP_t_full} also determines the translational symmetry of the system: the unit cell of the mobile-ion sublattice should correspond to the least common multiple of the spatial periodicity of the dominant contributing modes. For example, if only the $\Gamma$-point modes are considered, this dynamic unit cell coincides with the static crystal unit cell. We then redefine $i,j$ as the unit cell index, with $a,b$ for inequivalent sites within the unit cell. Each site $(i,a)$ is at the position $\bm{r}_{ia}$. The hopping matrix is then denoted as $P_{ji}(t)\rightarrow P_{jb,ia}(t)$, and the ionic density for each unit cell is expressed as the vector $n_j(t) = (n_{j,a_1}(t), n_{j,a_2}(t), \ldots, n_{j,a_{N_u}}(t))^{\bm{T}}$, with Fourier transformation formed as: 
\begin{equation}
    \begin{aligned}
        n_{\bm{k},a}(t) &= \frac{1}{\sqrt{N_s}}{\sum_j^{N_s}}e^{-i\bm{k}\cdot\bm{r}_{ja}}n_{j,a}(t),
    \end{aligned}
\end{equation}
which leads Eq.\,\eqref{eq:njt_Pjit} into the reciprocal space expression: 
\begin{equation}
    \begin{aligned}
        &n_{\bm{k},a}(t+\tau) 
        = 
        {\sum_{\bm{k}'}^{N_s}} \left(  \frac{1}{N_s}{\sum_{i}^{N_s}}e^{i(\bm{k}'-\bm{k})\cdot\bm{r}_{ia}} \right) 
        \\
        &{\times}
        \left( {\sum\limits_{jb}^{N_s}}e^{-i\bm{k}\cdot(\bm{r}_{jb}-\bm{r}_{ia})}P_{jb,ia}(t) \right) n_{\bm{k}'}(t). 
    \end{aligned}
\end{equation}
Here the reciprocal-space ionic density vector is defined as $n_{\bm{k}}(t) = (n_{\bm{k},a_1}(t),n_{\bm{k},a_2}(t),\ldots,n_{\bm{k},a_{N_u}}(t))^{\bm{T}}$. Consider the translational symmetry such that the local hopping environment is identical for all the unit cells. In this case, the Fourier component of the hopping probability defined as 
\begin{equation}\label{eq:P_k_raw}
    P_{\bm{k},ba}(t) = \sum_{j}^{N_s} e^{-i\bm{k}\cdot(\bm{r}_{jb} - \bm{r}_{ia})} {P}_{jb,ia}(t)
\end{equation}
is independent of $i$. We further denote $\mathbb{P}_{\bm{k}}(t)$ as the $(N_u\times N_u)$-dimensional hopping matrix on the basis of inequivalent sites, i.e. $[\mathbb{P}_{\bm{k}}(t)]_{ba} = P_{\bm{k},ba}(t)$. This symmetry allows the conservation of $\bm{k}$ as $\frac{1}{N_s} \sum_{i=1}^{N_s} e^{i(\bm{k}' - \bm{k})\cdot\bm{r}_{ia}} = \delta_{\bm{kk}'}$, and the dynamic equation becomes diagonal in $k$-space as:
\begin{equation}\label{eq:nPn_t_ink}
    n_{\bm{k}}(t+\tau) = \mathbb{P}_{\bm{k}}(t)n_{\bm{k}}(t).
\end{equation}
At thermodynamic equilibrium, the hopping energy dissipates to the host lattice, allowing the near-neighbor approximation that only hops within a given cutoff distance are considered. 

We further consider the normalized eigenvectors ${u}_{\bm{k},\nu}(t)=(u_{\bm{k},\nu,a_1}(t),u_{\bm{k},\nu,a_2}(t),\ldots,u_{\bm{k},\nu,a_{N_u}}(t))^{\bm{T}}$ and eigenvalues $p_{\bm{k},\nu}$ of $\mathbb{P}_{\bm{k}}(t)$ that satisfy: 
\begin{equation}
    \begin{aligned}
        \mathbb{P}_{\bm{k}}(t){u}_{\bm{k},\nu}(t) &= p_{\bm{k},\nu}(t){u}_{\bm{k},\nu}(t), 
    \end{aligned}
\end{equation}
where $\nu$ is the eigenmode index. For these eigenmodes, Eq.\,\eqref{eq:nPn_t_ink} becomes ${u}_{\bm{k},\nu}(t+\tau) = p_{\bm{k},\nu}(t){u}_{\bm{k},\nu}(t)$. Therefore, for a period $N\tau$ the temporal evolution becomes: 
\begin{equation}\label{eq:u_k_nu_t_Ntau}
    {u}_{\bm{k},\nu}(t+N\tau) = \left( \prod_{l=1}^N |p_{\bm{k},\nu}(t_l)|e^{i\tau\omega_{\bm{k},\nu}(t_l)} \right) {u}_{\bm{k},\nu}(t),
\end{equation}
where $t_l=t+l\tau,l=1,\ldots,N$. It is clear that the magnitude of the eigenvalue determines stability: a mode is stable only when $\Pi_{l=1}^N|p_{\bm{k},\nu}(t_l)|=1$, or quasi-stable when $\Pi_{l=1}^N|p_{\bm{k},\nu}(t_l)|\sim1$. The characteristic decay rate is given as: 
\[ \kappa_{\bm{k},\nu}(t)\equiv (1 - |p_{\bm{k},\nu}(t)|)/\tau. \]
On the other hand, the argument angle $\arg[p_{\bm{k},\nu}(t)]$ is associated with the frequency $\omega_{\bm{k},\nu}(t)$ as: 
\[ \omega_{\bm{k},\nu}(t) \equiv \arg[p_{\bm{k},\nu}(t)]/\tau. \]

The eigenmodes $\{ {u}_{\bm{k},\nu}(t) | \nu=1,2,\ldots,N_u \}$ form a complete basis that can express any ionic density in real space through the inverse Fourier transformation as: 
\begin{equation}\label{eq:n_j_raw}
    \begin{aligned}
        {n}_{j,b}(t) &= {\rm Re}[\frac{1}{\sqrt{N_s}}{\sum\limits_{\bm{k},\nu}}e^{i\bm{k}\cdot\bm{r}_{jb}} {n}_{\bm{k},\nu,b}(t)], 
    \end{aligned}
\end{equation}
where ${n}_{\bm{k},\nu,b}(t)\equiv c_{\bm{k},\nu,b}(t)u_{\bm{k},\nu,b}(t)$ is the projection of the ionic density onto the eigenmode $\nu$ at the wave-vector $\bm{k}$, with $c_{\bm{k},\nu,b}(t)$ the corresponding amplitude. Note that only the real component is needed in Eq.\,\eqref{eq:n_j_raw} since the ionic density is a real-valued classical quantity, and the imaginary part of $\sum_{\bm{k},\nu}e^{i\bm{k}\cdot\bm{r}_{jb}}{n}_{\bm{k},\nu,b}(t)$ is just introduced for convenience in the Fourier transformation. 

We can further evaluate the local ionic current density through the ionic polarization $Z \bm{r}_{jb} n_{j,b}(t)$ as: 
\begin{equation}\label{eq:J_full_raw}
    \begin{aligned}
        \bm{J}^{\rm full}_{jb}(t) 
        &= -\frac{\partial}{\partial t}\!\left[ Z \bm{r}_{jb} n_{jb}(t) \right] \\
        &= -Z \frac{\partial}{\partial t}\frac{1}{\sqrt{N_s}} 
        {\rm Re}\!\left[\sum_{\bm{k},\nu} 
        \bm{r}_{jb} e^{i\bm{k}\cdot\bm{r}_{jb}} n_{\bm{k},\nu,b}(t)\right]
        \\
        &= -iZ \frac{\partial}{\partial t} \frac{1}{\sqrt{N_s}}
        {\rm Re}\!\left[\sum_{\bm{k},\nu} 
        e^{i\bm{k}\cdot\bm{r}_{jb}} \nabla_{\bm{k}} n_{\bm{k},\nu,b}(t)\right],
    \end{aligned}
\end{equation}
where $Z$ is the effective ionic charge. Note that although the polarization is only well-defined with respect to a chosen reference position~\cite{vanderbilt1993electric}, this ambiguity does not affect the definition of the current as it only involves \textit{changes} in polarization. 

Up to this point, apart from assuming translational symmetry of the crystalline lattice and one single representative dynamic scale $\tau$, only approximations (1) and (3) introduced in Sec.\,\ref{sec:approximations} have been applied.



\subsection{Lattice model for normal solid-state ionic conductors}\label{sec:lattice_model_conventional}

Starting from the general framework above, we can further adopt approximations (2) and (4) of Sec.\,\ref{sec:approximations} to recover the standard theory for normal solid-state ionic conductors. Specifically, by assuming an adiabatic potential energy landscape and negligible interactions between mobile ions, the hopping matrix $\mathbb{P}_{\bm{k}}(t)\rightarrow \mathbb{P}_{\bm{k}}$ is no longer time dependent (in a static external field), nor are $\kappa_{\bm{k},\nu}$ and $\omega_{\bm{k},\nu}$. In this case, Eq.\,\eqref{eq:u_k_nu_t_Ntau} simplifies to: 
\begin{equation}
    u_{\bm{k},\nu}(t+N\tau) = (1- \kappa_{\bm{k},\nu}\tau)^Ne^{iN\tau\omega_{\bm{k},\nu}} u_{\bm{k},\nu}(t). 
\end{equation}
For stable modes we simply have $n_{\bm{k},\nu}(t) = n_{\bm{k},\nu}(0)e^{i\omega_{\bm{k},\nu}t}$. Correspondingly, according to Eq.\,\eqref{eq:n_j_raw}, any stable ionic density distribution is in the form: 
\begin{equation}
    \begin{aligned}
        n_{jb}(t) &= 
        \frac{1}{\sqrt{N_s}}{\sum\limits_{\bm{k},\nu}}|n_{\bm{k},\nu,b}(0)|\cos(\bm{k}\cdot\bm{r}_{jb} + \omega_{\bm{k},\nu}t + \phi_{\bm{k},\nu,b}), 
    \end{aligned}
\end{equation}
where $\phi_{\bm{k},\nu,b}$ and $|n_{\bm{k},\nu,b}(0)|$ are the initial phase and strength respectively. In this case, Eq.\,\eqref{eq:J_full_raw} gives a macroscopically homogeneous component of the drift ionic current as follows: 
\begin{equation}
    \begin{aligned}
        \bm{J}(t) &= {\sum\limits_{\nu,a}}Zn_{\bm{0},\nu,a}(t)\bm{v}_{\bm{0},\nu}, 
    \end{aligned}
\end{equation}
where the group velocity $\bm{v}_{\bm{k},\nu} \equiv \nabla_{\bm{k}} \omega_{\bm{k},\nu}$ refers to the velocity of the ionic density wave packet. Correspondingly, the macroscopic ionic conductivity is calculated as: 
\begin{equation}
    \begin{aligned}
        \sigma_{\mu\alpha}(t) = \frac{\partial }{\partial \mathcal{E}^{\alpha}}J^{\mu}(t), 
    \end{aligned}
\end{equation}
where $\mu,\alpha$ refer to the spatial direction, and $\mathcal{E}^{\alpha}$ is the $\alpha$-direction electric field intensity. Specifically, if $\omega_{\bm{0},\nu}=0$ so that $n_{\bm{0},\nu,a}(t)\rightarrow n_{\bm{0},\nu,a}$ becomes a constant, we obtain the following DC ionic conductivity: 
\begin{equation}\label{eq:sigma_dc_gamma}
    \begin{aligned}
        \sigma_{\mu\alpha} = {\sum\limits_{\nu,a}}Zn_{\bm{0},\nu,a} \frac{\partial v^{\mu}_{\bm{0},\nu}}{\partial \mathcal{E}^{\alpha}} 
        + 
        {\sum\limits_{\nu,a}}Z\frac{\partial n_{\bm{0},\nu,a}}{\partial \mathcal{E}^{\alpha}}v^{\mu}_{\bm{0},\nu}, 
    \end{aligned}
\end{equation}
where ${\partial v^{\mu}_{\bm{0},\nu}}/{\partial \mathcal{E}^{\alpha}}$ is the mode-resolved ionic mobility.


\section{Benchmark for normal ionic conductors}

In this section, we apply the reduced theory given in Sec.\,\ref{sec:lattice_model_conventional} to a one-dimensional, two-site model, as a benchmark for normal ionic conductors. 

\subsection{Intrinsic ionic conduction}\label{sec:intrinsic_1D}

\begin{figure*}[htp]
    \centering
    \includegraphics[width=0.7\textwidth]{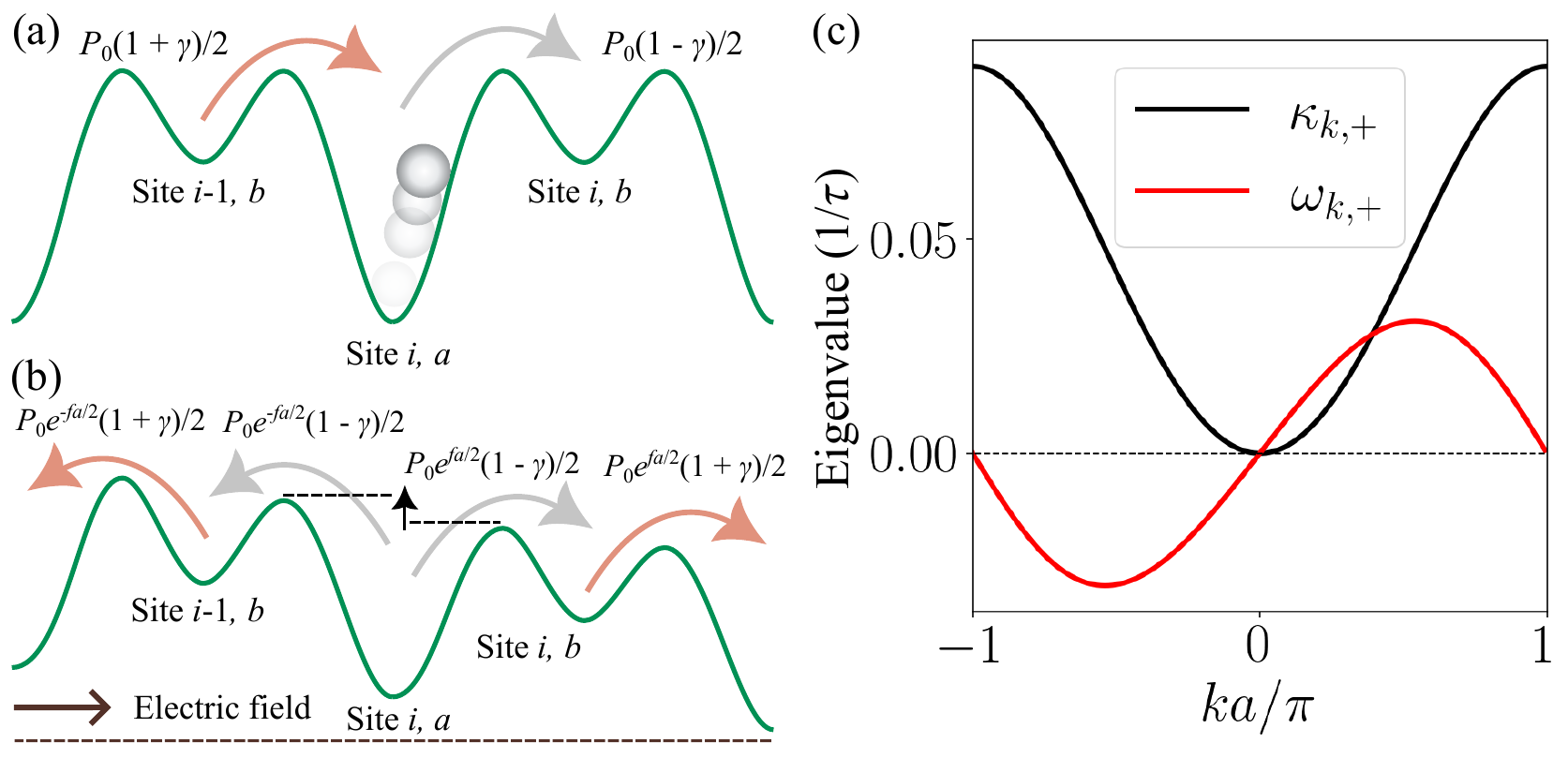}
    \caption{\label{fig:schematic_1D}
    \textbf{Results of 1D model.} Schematic (a) without and (b) with the external electric field. 
    (c) Eigenvalues of `$+$' branch with parameters $P_0=0.60, \gamma=0.88,f=0.5$. The absolute stable mode is at $k=0$ with $\kappa_{k,+}=0$.}
\end{figure*}


Consider a simple but representative one-dimensional example with two inequivalent sites, $a$ and $b$, in each unit cell, where each $a$ site is surrounded by two $b$ sites and vice versa. The single-particle, adiabatic hopping probability from $a$ to $b$ is $P_{ba} = P_0(1-\gamma)/2$, where the factor $1/2$ accounts for site degeneracy, and $-1<\gamma<1$ arises from the staggered potential landscape, as schematized in Fig.\,\ref{fig:schematic_1D}(a). The probability of hopping in reverse is $P_{ab}=P_0(1+\gamma)/2$, which suggests the following relations: 
\begin{equation}\label{eq:P0_gamma_definition}
    P_0 = P_{ab} + P_{ba}, \ \ \frac{1-\gamma}{1+\gamma}= \frac{P_{ba}}{P_{ab}}. 
\end{equation}
The nearest-neighboring hopping probability in real space becomes: 
\begin{equation}
    \begin{aligned}
        P_{ib,ia} &= \frac{P_0(1-\gamma)}{2} = P_{(i-1)b,ia}, 
        \\
        P_{ia,ib} &= \frac{P_0(1+\gamma)}{2} = P_{(i+1)a,ib}, 
        \\
        P_{ia,ia} &= 1 - P_0(1-\gamma),
        \ \
        P_{ib,ib} = 1 - P_0(1+\gamma),
    \end{aligned}
\end{equation}
and is zero for other cases. 

In reciprocal space with the basis $(n_{k,a}, n_{k,b})^{\bm{T}}$, the hopping matrix associated with mode $k$ follows Eq.\,\eqref{eq:P_k_raw} and is given by: 
\begin{equation}
    \mathbb{P}_k =
    \begin{bmatrix}
        1 - P_0(1 - \gamma) & P_0(1+\gamma)\cos(ka/2) \\
        P_0(1-\gamma)\cos(ka/2) & 1 - P_0(1+\gamma)
    \end{bmatrix}.
\end{equation}
The eigenvalues for the two modes $\nu=+,-$ are: 
\begin{equation}
    p_{k,\pm} = 1 - P_0 \pm P_0 \sqrt{\cos^2(ka/2) + \gamma^2 \sin^2(ka/2)},
\end{equation}
which are purely real, leading to no propagation of the ionic density wave. Only the `$+$' mode at the $\Gamma$ point is absolutely stable with $|p_{k,+}|=1$, as expected in the absence of interactions breaking the translational symmetry. For small $k$, the decay rate behaves as: 
\begin{equation}\label{eq:kappa_k0_1D}
    \kappa_{k,+} = \frac{1-|p_{k,+}|}{\tau} = \frac{P_0}{\tau} \frac{1}{2} (1-\gamma^2) \left(\frac{ka}{2}\right)^2 + o[(ka)^4].
\end{equation}
This result indicates that any macroscopically inhomogeneous ionic density pattern has a finite lifetime, and that it scales with the wave-vector $k$ as $\kappa_k \propto k^2$, which is consistent with the diffusion process. 

The eigenvector $u_{k,+} = (u_{k,+,a}, u_{k,+,b})^{\bm{T}}$ satisfies $\mathbb{P}_ku_{k,+} = p_{k,+}u_{k,+}$. At $k=0$, the ionic density ratio between the two sites recovers the hopping probability ratio as:  
\begin{equation}\label{eq:nbna_1D}
    \frac{u_{0,+,b}}{u_{0,+,a}} = \frac{1-\gamma}{1+\gamma} = \frac{P_{ba}}{P_{ab}}, 
\end{equation}
which is a result of detailed balance between the two sites. In the long-wavelength limit $k \to 0$, we obtain the microscopic polarization: 
\begin{equation}\label{eq:polarization_1D}
    \begin{aligned}
        \frac{u_{k,+,a} - u_{k,+,b}}{u_{k,+,a} + u_{k,+,b}}
        = \gamma + \gamma(1 - \gamma^2)\left(\frac{ka}{2}\right)^2 + o[(ka)^4], 
    \end{aligned}
\end{equation}
which results from the potential landscape bias $\gamma$. The first term $\gamma$ is the homogeneous polarization, while the $k$-dependent term relates to the domain structure.


\subsection{Coupling with electric field}\label{sec:couple_E_1D}

In this section, we consider a homogeneous electric field $\mathcal{E}$ applied to the system. In the simple-hopping case, as given in Eq.\,\eqref{eq:P_ji_simple}, the hopping barrier is shifted by $\pm fa/2$ for opposite directions with $f\equiv Z\mathcal{E}/k_{\rm B}T$. As a result, the hopping probabilities are modified as: 
\begin{equation}
    \begin{aligned}
        P_{ib,ia} &= \frac{P_0(1-\gamma)}{2}e^{fa/2}, \ \  P_{(i-1)b,ia} = \frac{P_0(1-\gamma)}{2}e^{-fa/2}, 
        \\
        P_{ia,ib} &= \frac{P_0(1+\gamma)}{2}e^{-fa/2}, \ \ P_{(i+1)a,ib} = \frac{P_0(1+\gamma)}{2}e^{fa/2}, 
        \\
        P_{ia,ia} &= 1 - P_0(1-\gamma)\cosh\left(\frac{fa}{2}\right),
        \\
        P_{ib,ib} &= 1 - P_0(1+\gamma)\cosh\left(\frac{fa}{2}\right). 
    \end{aligned}
\end{equation}
If the density of states $g(E)$ in Eq.\,\eqref{eq:P_ji_DOS} can be significantly modified by $\mathcal{E}$ (e.g. through structure distortion or electronic polarization), a screening factor should be included, and in the following discussion we assume $\mathcal{E}$ and $f$ correspond to their screened values, leaving detailed discussion to future work. Therefore, in reciprocal space we have: 
\begin{equation}\label{eq:Pk_1D_Efield}
    \begin{aligned}
        \mathbb{P}_k &= 
        \begin{bmatrix}
            1-P_0(1 - \gamma)\Omega(0, -f) & P_0(1+\gamma)\Omega(k,f) \\
            P_0(1-\gamma)\Omega(-k,-f) & 1-P_0(1+\gamma)\Omega(0, f) 
        \end{bmatrix}, 
    \end{aligned}
\end{equation}
where the field-modified structural factor is given as:
\begin{equation}\label{eq:Omega_k_f_1D}
    \begin{aligned}
        &\Omega(k, f) \equiv \frac{1}{2}(e^{-ika/2 - fa/2} + e^{ika/2 + fa/2}) 
        \\
        &= \cos\left(\frac{ka}{2}\right)\cosh\left(\frac{fa}{2}\right) + i\sin\left(\frac{ka}{2}\right)\sinh\left(\frac{fa}{2}\right),
    \end{aligned}
\end{equation}
In this case the eigenvalues become: 
\begin{equation}\label{eq:p_k_pm_E_1D}
    \begin{aligned}
        p_{k,\pm} &= 1 - P_0\Omega(0,f)\left( 1 \mp \sqrt{\gamma^2 + (1-\gamma^2)[\frac{\Omega(k,f)}{\Omega(0,f)}]^2} \right), 
    \end{aligned}
\end{equation}
which are now complex numbers, and, as discussed in Eq.\,\eqref{eq:u_k_nu_t_Ntau}, suggests a non-zero frequency of ionic density wave. 

The only stable mode is in the `$+$' branch at the $\Gamma$ point, satisfying $|p_{0,+}|=1$. The full spectrum with representative parameters is shown in Fig.\,\ref{fig:schematic_1D}(c). In the weak-field limit $|fa| = |\frac{Za\mathcal{E}}{k_{\rm B}T}|\ll 1$ and near the $\Gamma$ point with $|ka|\ll1$, we have: 
\begin{equation}
    p_{k,+} \approx 1 + iP_0(1-\gamma^2)\frac{ka}{4}fa,
\end{equation}
which leads to a non-zero wave frequency: 
\begin{equation}
    \omega_{0,+} = \frac{a}{\tau}P_0(1-\gamma^2)\left(\frac{aZ\mathcal{E}}{4k_{\rm B}T}\right).
\end{equation}
In contrast, the changes that an electric field introduces to the decay rate, that is, to the magnitude of $p_{k,+}$, are in the second-order. Therefore the key consequence of a weak electric field is making the ionic density wave propagate. The group velocity $v_{0,+}$ is calculated as: 
\begin{equation}
    \begin{aligned}
        v_{0,+}(\mathcal{E}) = \frac{a}{\tau}P_0(1-\gamma^2)\left(\frac{aZ\mathcal{E}}{4k_{\rm B}T}\right)
    \end{aligned}
\end{equation}
which, as aforementioned, refers to the velocity of the ionic density wave packet. The eigenstate associated with the `$+$' mode $u_{k,+}=(u_{k,+,a},u_{k,+,b})^{\bm{T}}$ is approximated as: 
\begin{equation}\label{eq:1_pigamma}
    \begin{aligned}
        &\frac{u_{k,+,b}}{u_{k,+,a}} \approx \left(\frac{1 - \gamma}{1+\gamma}\right)\left( 1 + i\gamma\frac{ka}{4}\frac{aZ\mathcal{E}}{k_{\rm B}T} \right), 
    \end{aligned}
\end{equation}
which indicates that the electric field brings a phase difference $\Delta \phi_k = \gamma\frac{ka}{4}\frac{aZ\mathcal{E}}{k_{\rm B}T}$ to the ionic density wave between the $a, b$ sites, while the ionic polarization (\textit{i.e.} the real component) is not significantly changed. 

According to Eq.\,\eqref{eq:sigma_dc_gamma}, a non-zero group velocity $v_{0,+}$ leads to a macroscopic DC ionic conductivity: 
\begin{equation}\label{eq:sigma_1D}
    \sigma_{\rm weak} = \frac{n_0Za}{V\tau}\frac{Za}{4k_{\rm B}T}(1-\gamma^2)P_0, 
\end{equation}
where $Z$ is the ionic charge, $n_0$ is the filling number of mobile ions within each unit cell and should not be larger than the number of inequivalent sites (i.e., $n_0 \le N_u = 2$ in this case). If we further assume the simple structure of local hopping $P_0=\exp(-\Lambda_0/k_{\rm B}T)$ as introduced in Eq.\,\eqref{eq:P_ji_simple}, and other parameters are insignificantly dependent on temperature $T$, we have:
\begin{equation}
    \ln(\sigma_{\rm weak} T) = -\frac{\Lambda_0}{k_{\rm B}T} + \rm{Const.}, 
\end{equation} 
which is the typical Arrhenius law for ionic conductivity of normal solid-state ionic conductors. 

In the presence of a strong field and if the stable mode is still near the $\Gamma$ point, the structural factor of Eq.\,\eqref{eq:Omega_k_f_1D} becomes: 
\begin{equation}
    \begin{aligned}
        \frac{ \Omega(k, f) }{ \Omega(0, f) } &= \frac{e^{-ika/2} + e^{ika/2}e^{fa} }{1 + e^{ fa}}
        \\
        &\rightarrow 1 + i\frac{ka}{2} \tanh\left(\frac{fa}{2}\right) + o[(ka)^2], 
    \end{aligned}
\end{equation}
and the wave frequency of mode `$+$' becomes: 
\begin{equation}
    \omega_{k,+} = P_0(1-\gamma^2) \sinh\left(\frac{fa}{2}\right)\frac{ka}{2}, 
\end{equation}
Therefore the macroscopic DC conductivity in the strong-field limit becomes:
\begin{equation}
    \sigma_{\rm strong} = \sigma_{\rm weak} \cosh\left(\frac{\mathcal{E}a}{2k_{\rm B}T}\right). 
\end{equation}
where $\sigma_{\rm weak}$ is the weak-field counterpart given in Eq.\,\eqref{eq:sigma_1D}. It agrees well with the empirical exponential law observed in experiments at high field~\cite{kemp2021nonlinear}. 

In summary, inter-site hopping under translational symmetry gives rise to a finite lifetime for inhomogeneous modes with $k \neq 0$, whereas only the homogeneous mode ($k = 0$) can persist over long times. These modes represent propagating ionic density waves driven by an external electric field: the $k = 0$ mode possesses a finite group velocity but zero frequency, while the $k \neq 0$ modes have both finite group velocity and finite frequency. Consequently, the ionic current reaches a steady state only after a sufficiently long timescale, when all inhomogeneous and oscillatory components have decayed. This result aligns with experimental observations showing that the system gradually loses its ``memory'' of the external driving force (e.g. optical impulsion) over a finite relaxation period~\cite{poletayev2024persistence}.


\begin{figure*}[htp]
    \centering
    \includegraphics[width=0.8\textwidth]{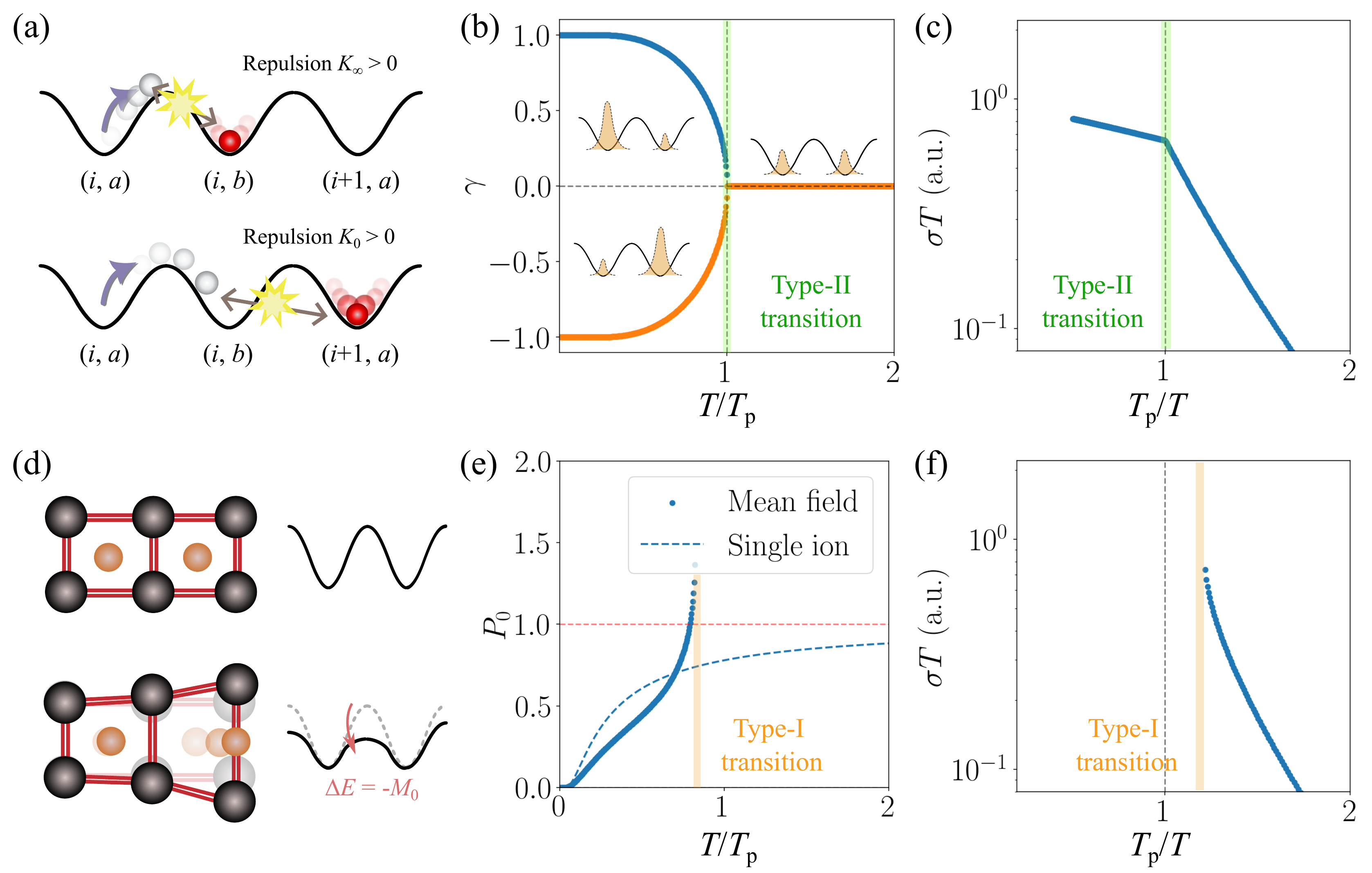}
    \caption{\label{fig:1D_results}
    \textbf{Self-consistent mean-field calculation of one-dimensional model.} (a) Schematic of many-body effect: the hopping ion will receive Coulomb repulsion from neighboring mobile ions. (b) Ionic polarization $\gamma$ and (c) ionic conductivity with weak nonadiabatic strength $M_0=0.25$~eV, which yield type-II superionic phase transition. (d) Schematic of nonadiabatic hopping: when one ion hops, it compresses the host lattice so that reduces the hopping barrier of the neighboring ions. (e) Average hopping possibility $P_0$, and (f) ionic conductivity with strong nonadiabatic effect $M_0=2.0$~eV, which yield type-I superionic phase transition. Data in type-I superionic phase is not plotted since they are not valid due to the breakdown of tight-binding approximation. Inter-site Coulomb repulsion strength $K_0=0.5$~eV; single-particle adiabatic hopping barrier $\Lambda_0=0.25$~eV; filling number $n_0=1$. Note that the type-II transition temperature $T_{\rm p}$ is used as a reference temperature for the plot.}
\end{figure*}

\section{Origins of superionic phase transitions}\label{sec:origins_of_superionic_phase_transition}

In this section, based on the one-dimensional model results of normal ionic conductors discussed above, we break the single-particle approximation and adiabatic approximation to study SICs. 

\subsection{Many-body effects as the polarization mean field}\label{sec:depol_meanfield_1D}

As schematicized in Fig.~\ref{fig:1D_results}
(a), we consider the many-body effect arising from the Coulomb interaction between mobile ions: (i) if a site is occupied by more than one ion, their energy increases drastically by $K_{\infty}$, as such double occupation is energetically forbidden; (ii) an ion hopping from site $(i,a)$ to site $(i,b)$ experiences a Coulomb repulsion energy $K_0$ from ions at the nearest neighboring site $(i+1,a)$. Therefore, the energy barrier of $(i,a)\rightarrow (i,b)$ hopping $E_{(i,b)(i,a)}$ is shifted by: 
\begin{equation}
    \Delta E_{(i,b)(i,a)} = K_{\infty}\,\theta(n_{i,b}-1) + K_0n_{i+1,a}, 
\end{equation}
where $\theta(x)$ is the step function, equal to $1$ for $x\geq 0$ and to $0$ otherwise. 

In principle, the occupation densities $n_{i,b}$ and $n_{i+1,a}$ are time dependent and can vary from different evolution trajectories. Here to capture lowest-order effects, we adopt the \textit{self-consistent mean-field method} by considering only the expectation value of the ionic density, while their transient components are neglected. This expectation value refers to the stable solution as mentioned in Eq.\,\eqref{eq:nbna_1D}, because whatever the initial state is, the system will converge to this stable state after a long time evolution. In this case, the ionic density is given as: 
\begin{equation}\label{eq:na_nb_k0}
    n_{i,a} \rightarrow n_0\frac{1}{2}(1+\gamma), 
    \quad 
    n_{i,b} \rightarrow n_0\frac{1}{2}(1-\gamma). 
\end{equation}
Hence we can drop the $i$-dependence due to the translational symmetry of the stable solution, and the energy barrier of $a\rightarrow b$ hopping is shifted by: 
\begin{equation}
    \begin{aligned}
        \Delta E_{ba} &= K_{\infty}\theta\left(n_0\frac{1-\gamma}{2}-1\right) + K_0n_0\frac{1}{2}(1+\gamma). 
    \end{aligned}
\end{equation}
Similarly, the hopping energy shift of the reverse process is: 
\begin{equation}
    \Delta E_{ab} = K_{\infty}\theta\left(n_0\frac{1+\gamma}{2}-1\right) + K_0n_0\frac{1}{2}(1-\gamma). 
\end{equation}
In this case, if the bias factor of the potential energy landscape $\gamma$ is totally driven by this many-body effect (i.e. $P_{ab}=P_{ba}=\exp(-\Lambda_0/k_{\rm B}T)/2$ without considering many-body effects), we have the hopping probabilities given as: 
\begin{equation}\label{eq:Pab_ba_gamma_meanfield}
    \begin{aligned}
        P_{ba} &= \frac{1}{2}\exp\left(-\frac{\Lambda_0 + K_{\infty}\theta(n_0\frac{1-\gamma}{2}-1) + K_0n_0(\frac{1+\gamma}{2})}{k_{\rm B}T}\right), \\
        P_{ab} &= \frac{1}{2}\exp\left(-\frac{\Lambda_0 + K_{\infty}\theta(n_0\frac{1+\gamma}{2}-1) + K_0n_0(\frac{1-\gamma}{2})}{k_{\rm B}T}\right). 
    \end{aligned}
\end{equation}
To eliminate the unphysical double-occupation term associated with $K_{\infty}$, the condition $|\gamma| < 2/n_0 - 1$ must be satisfied. For example, when $n_0 = 1$, double occupation is completely excluded within the mean-field framework, imposing no additional constraint on $\gamma$ beyond its original definition $|\gamma| < 1$ (given in Eq.\,\eqref{eq:P0_gamma_definition}). 

On the other hand, we have the original definition of $P_{ab}, P_{ba}$ given in Eq.\,\eqref{eq:P0_gamma_definition} as: 
\begin{equation}
    \begin{aligned}
        P_{ba} &= P_0(1-\gamma)/2, \ \ \ 
        P_{ab} = P_0(1+\gamma)/2,
    \end{aligned}
\end{equation}
which, as a requirement of self-consistency, should be satisfied simultaneously with Eq.\,\eqref{eq:Pab_ba_gamma_meanfield}. Therefore, we find the relation between $P_0$ and $\gamma$ as: 
\begin{equation}\label{eq:P0_gamma_1meanfield}
    \begin{aligned}
        P_0 &= \exp\left(-\frac{\Lambda_0 + n_0K_0/2}{k_{\rm B}T}\right)\cosh\left(\frac{n_0K_0\gamma}{2k_{\rm B}T}\right), 
    \end{aligned}
\end{equation}
and, most importantly, the self-consistency equation for $\gamma$ as:
\begin{equation}\label{eq:gamma_selfconsistent}
    \gamma = \tanh\!\left(\frac{\gamma n_0K_0}{2k_{\rm B} T}\right).
\end{equation} 
This equation defines the phase diagram of $\gamma$ as a function of temperature $T$ for a given filling number $n_0$ and inter-site Coulomb repulsion strength $K_0$. According to Eq.\,\eqref{eq:polarization_1D}, $\gamma$ is associated with the permanent ionic polarization, hence we name this mean field the `polarization mean field'. The critical temperature is simply given by: 
\begin{equation}\label{eq:Tp_1D}
    T_{\rm p} = \frac{n_0K_0}{2k_{\rm B}}.
\end{equation}
The calculated phase diagram of $\gamma$ with $n_0=1$ is shown in Fig.\,\ref{fig:1D_results}(b). For $T<T_{\rm p}$, two solutions with opposite polarizations $\pm \gamma$ exist, while for $T>T_{\rm p}$ the ionic system is depolarized with $\gamma=0$. The transition is second-order as illustrated by the continuous but non-smooth feature at $T_{\rm p}$. Using Eq.\,\eqref{eq:sigma_1D}, we can construct the phase diagram for ionic conductivity, as shown in Fig.\,\ref{fig:1D_results}(c). 

These results recover the following experimental observations of type-II superionic phase transitions: (i) the temperature dependence of the conductivity shows a second-order transition with different Arrhenius fitting in the two phases~\cite{BOUKAMP19831193,boyce1979superionic,krenzer2023nature,wang2021atomistic}; (ii) the polarization $\gamma$ decreases with increasing temperature until it vanishes, which agrees well with X-ray spectrum measurements of the structure factor transition and with molecular dynamics simulations~\cite{pnas}; (iii) the transition temperature $T_{\rm p}$ is only determined by the inter-site Coulomb interaction $K_0$ and filling number $n_0$, and does not depend on single-ion properties and other chemical details, which can be directly verified by the fact that CuCrS$_2$ and AgCrS$_2$ and their alloys show almost the same transition temperature~\cite{yakshibaev1991ionic}; (iv) the mobile ions remain solid-like in the superionic phase, which is consistent with the experimental observation of long-wavelength transverse acoustic phonons persisting across the superionic transition~\cite{ding2025atomic}.


\subsection{Nonadiabatic effects as the concerted-hopping mean field}\label{sec:concert_meanfield_1D}

We further consider nonadiabatic effects whereby the ion at site $(i,a)$ can be influenced by hopping events occurring at its nearest-neighboring sites. These effects arise from two mechanisms: (i) as illustrated schematically in Fig.\,\ref{fig:1D_results}(d), when an ion hops from $(i-1,b)$ to $(i-1,a)$ (or from $(i,b)$ to $(i+1,a)$), it locally compresses the host lattice and thereby creates additional space for the ion at $(i,a)$ to hop toward $(i-1,b)$ (or $(i,b)$), effectively reducing the corresponding hopping barrier; (ii) conversely, when an ion hops from $(i-1,a)$ to $(i-1,b)$ (or from $(i+1,a)$ to $(i,b)$), part of its hopping energy can be transferred to the ion at $(i,a)$ through Coulomb repulsion, which also effectively lowers the barrier for the $(i,a)\!\rightarrow\!(i,b)$ (or $(i,a)\!\rightarrow\!(i-1,b)$) hopping process.  

In particular, if the dynamics of the host lattice are much faster than those of the mobile-ion hops, the lattice distortion associated with mechanism (i) will rapidly relax, and the transferred hopping energy in mechanism (ii) will be quickly dissipated into the host lattice vibrations. As a result, these effects become negligible for subsequent hopping events. Therefore, this represents a typical nonadiabatic effect that becomes significant only when the dynamical timescales of the mobile-ion hopping and immobile lattice components are comparable.

To evaluate this nonadiabatic hopping barrier reduction, we have that the probability of one ion hopping from $(i-1,b)$ to $(i-1,a)$ or from $(i,b)$ to $(i+1,a)$ is $P_0(1+\gamma)/2$. Therefore the number of hopping ions is $n_{i-1,b}P_0(1+\gamma)/2$ or $n_{i,b}P_0(1+\gamma)/2$. We define the lowest-order nonadiabatic strength as $M_0>0$, and the hopping barrier is reduced by: 
\begin{equation}
    \begin{aligned}
        \Delta E_{(i,b)(i,a)} &= -M_0n_{i,b}P_0\frac{1+\gamma}{2}, \\ 
        \Delta E_{(i-1,b)(i,a)} &= -M_0n_{i-1,b}P_0\frac{1+\gamma}{2}.
    \end{aligned}
\end{equation}
We again adopt the mean-field approximation through Eq.\,\eqref{eq:na_nb_k0} to obtain: 
\begin{equation}
    \begin{aligned}
        \Delta E_{ba} &= -M_0n_0P_0\left(\frac{1-\gamma^2}{4}\right) = \Delta E_{ab}. 
    \end{aligned}
\end{equation}
Therefore, the concerted hopping does not affect $\gamma$ but only relates to $P_0$. Letting $\Lambda_0$ denote the single-particle adiabatic activation energy, the self-consistency equation for $P_0$ becomes: 
\begin{equation}
    P_0 = \exp\!\left[-\frac{\Lambda_0 - P_0n_0M_0 (\frac{1-\gamma^2}{4})}{k_{\rm B} T}\right].
\end{equation}
A typical phase diagram of $P_0$ is shown in Fig.\,\ref{fig:1D_results}(e), where $P_0$ diverges as it approaches a certain temperature, beyond which there is no self-consistency solution, indicating a drastic activation of the mobile ion sublattice. As shown in Fig.\,\ref{fig:1D_results}(f), the calculated conductivity presents a discontinuous (first-order) phase transition, which matches well with the experimental observation of a type-I superionic phase transition. 

This result suggests that a type-I phase transition induced by nonadiabatic effects corresponds to the \textit{melting} of the mobile-ion sublattice. Above the transition temperature, mobile ions are no longer primarily localized at discrete lattice sites but can significantly occupy high-energy regions as they actively hop between sites. Since this liquid-like behavior can be driven by the strong coupling between mobile and immobile ions, a type-I transition is usually accompanied by a significant structural change in the host lattice. This interpretation is consistent with simulations of type-I SIC Ag$_2$S~\cite{simonnin2020phase} and AgI~\cite{wood2006dynamical,hajibabaei2025symmetry}, which reveal that Ag$^+$ ions diffuse in a liquid-like manner without preferred diffusion pathways within the relatively stable body-centered cubic sulfur or iodine frameworks. It is also observed in a wide range of thermoelectric materials, including Cu$_{2-x}$Se~\cite{liu2012copper}, Zn$_4$Sb$_3$~\cite{snyder2004disordered,chalfin2007cation}, and Ag$_8$SnSe$_6$~\cite{ren2023extreme}.


\section{Discussion}

\subsection{Relationship between the two types of superionic phase transition}\label{sec:relation_meanfields_1D}

\begin{table*}[htbp]
    \centering
    \renewcommand{\arraystretch}{1.2}
    \begin{tabular}{ccc}
    \hline\hline
    \textbf{Category} & \textbf{Type-I} & \textbf{Type-II} \\
    \hline
    \textbf{Order of transition} 
        & First-order 
        & Second-order \\
    \textbf{Origin} 
        & Nonadiabatic effect
        & Many-body effect  \\
    \textbf{Type of mean field} 
        & Concerted hopping 
        & Polarization--depolarization \\
    \textbf{Type of self-consistency} 
        & Positive feedback 
        & Negative feedback \\
    \textbf{Polarity of mean field} 
        & Attractive interaction 
        & Repulsive interaction \\
    \textbf{Order parameter} 
        & Average hopping probability $P_0$ 
        & Polarization $\gamma$ \\
    \textbf{Role of increasing temperature} 
        & Activates mobile-ion sublattice 
        & Depolarizes mobile-ion sublattice \\
    \textbf{Physical picture} 
        & Melting of mobile-ion sublattice 
        & Depolarization of site occupation \\
    \textbf{Critical temperature} 
        & Depends on $\Lambda_0$, $M_0$, $n_0$, $K_0$ 
        & $T_{\rm p} \sim n_0 K_0 / k_{\rm B}$ \\
    \textbf{Dependence on Coulomb repulsion} 
        & Subtle 
        & Strong repulsion favors high $T_{\rm p}$ \\
    \textbf{Dependence on filling number} 
        & High value required 
        & Subtle \\
    \textbf{Dependence on host lattice} 
        & Strong coupling required 
        & Not sensitive \\
    \textbf{Experimental examples} 
        & AgX (X = Cl, Br, I, S$_{0.5}$) 
        & MCrX$_2$ (M = Cu, Ag; X = S, Se, Cl, Br, I) \\
    \hline\hline
    \end{tabular}
    \caption{\label{tab:comparison} \textbf{Comparison between two types of superionic phase transitions.} 
    The table summarizes their physical origin, self-consistency mechanism, order parameters, and characteristic material dependence.}
\end{table*}

As presented in Secs.~\ref{sec:depol_meanfield_1D} and~\ref{sec:concert_meanfield_1D}, the many-body and nonadiabatic effects can be effectively interpreted as polarization and concerted-hopping mean fields, which naturally give rise to type-II and type-I superionic phase transitions, respectively. Although the two types of superionic phase transition are not mutually exclusive, their coexistence is rarely observed due to their distinct physical mechanisms. Specifically, type-II and type-I transitions are effectively driven by repulsive and attractive interactions among mobile ions, respectively, representing two complementary aspects of collective ionic motion. As temperature increases, thermal activation enables the mobile-ion sublattice to overcome repulsive interactions and depolarize the site occupation, leading to a type-II transition. In contrast, for a type-I transition, sufficient thermal excitation allows mobile ions to efficiently utilize the hopping energy rather than dissipating it into the host lattice, giving rise to concerted hopping. This fundamental difference in physical origin leads to distinct dependencies of the two transition types on material properties:

\begin{enumerate}
    \item \textbf{Dependence on inter-ion Coulomb repulsion $K_0$:}  
    According to Eq.\,\eqref{eq:P0_gamma_2meanfield}, an experimentally accessible type-I transition temperature (at least below the melting point of the material) requires a small value of $\Lambda_0 + n_0 K_0/2$ and a large value of $n_0 M_0$. The former condition favors weak Coulomb repulsion, whereas the latter, as discussed in Sec.\,\ref{sec:concert_meanfield_1D}, requires strong Coulomb interaction. Consequently, realizing a type-I transition demands a delicate balance in the inter-ion Coulomb repulsion strength $K_0$, which depends on the screening efficiency, ionic species, and inter-site distance. This subtle balance explains why, although many concerted-hopping SICs exist, only a few exhibit measurable type-I superionic phase transitions.  
    In contrast, the transition temperature for type-II systems, given by Eq.\,\eqref{eq:Tp_1D}, is simply proportional to $K_0$, indicating that weaker Coulomb repulsion leads to a lower transition temperature.
    
    \item \textbf{Dependence on filling number $n_0$:}  
    Eq.\,\eqref{eq:Tp_1D} shows that a lower filling number suppresses the type-II transition temperature $T_{\rm p}$. Indeed, the dilute limit may be viewed as an extreme depolarized ``superionic phase,'' in which Coulomb repulsion is negligible and $T_{\rm p}$ becomes so low that conductivity in the normal phase is essentially unmeasurable. Moreover, because the type-II transition is a depolarization process, a sufficient number of vacant sites must be available at low temperature to form a polarized occupation pattern. In our two-inequivalent-site model, this requirement is encoded in the constraint $|\gamma| < 2/n_0 - 1$, as discussed in Eq.\,\eqref{eq:Pab_ba_gamma_meanfield}.  
    In contrast, type-I transitions generally favor higher filling numbers to enhance the concerted effect through $n_0 M_0$ in Eq.\,\eqref{eq:P0_gamma_2meanfield}. 
    
    \item \textbf{Dependence on host lattice:}  
    The type-I transition requires strong coupling between the host lattice and the mobile ions, such that the former is significantly modulated by the hopping dynamics of the latter. Thus, a ``softer'' host lattice is generally favored, as it corresponds to a lower single-ion adiabatic hopping barrier $\Lambda_0$~\cite{krauskopf2018comparing} and an enhanced coupling strength $M_0$. By contrast, the type-II transition is intrinsically insensitive to the microscopic details of the host lattice.
\end{enumerate}

These distinctions naturally lead each transition type to favor different classes of materials. Bulk compounds such as MX (M = Cu, Ag; X = Cl, Br, I, S$_{0.5}$) with wurtzite or hexagonal structures are more likely to exhibit type-I superionic phase transitions due to their compact lattice geometries. Conversely, layered materials such as MCrX$_2$ (M = Cu, Ag; X = S, Se), which feature a two-dimensional honeycomb lattice with half occupancy of mobile-ion sites, are more likely to undergo type-II transitions.

In addition to clarifying the differences between these two types of superionic transitions, our unified theoretical framework allows us to explore the possible \textit{coexistence} of both types of superionic phase transitions. 
Combining them yields a self-consistency equation for $P_0$:
\begin{equation}\label{eq:P0_gamma_2meanfield}
    \begin{aligned}
        P_0 &= \exp\!\left[-\frac{\Lambda_0 + n_0K_0/2 - P_0n_0M_0(1-\gamma^2)/4}{k_{\rm B}T}\right]
        \\
        &{\times}
        \cosh\!\left(\frac{n_0K_0\gamma}{2k_{\rm B}T}\right)
    \end{aligned},
\end{equation}
while the self-consistency equation for $\gamma$ remains unchanged and is still given by Eq.\,\eqref{eq:gamma_selfconsistent}. It follows that the transition of the concerted-hopping mean-field can only occur in the fully or partially depolarized regime, where $\gamma^2$ is small. Therefore, if both types of superionic phase transitions can really coexist in a SIC, the type-I transition must occur at a higher temperature than the type-II transition. 

Furthermore, it is important to emphasize that although the results in Secs.~\ref{sec:depol_meanfield_1D} and~\ref{sec:concert_meanfield_1D} are derived from a one-dimensional model, the central conclusions remain applicable to higher-dimensional systems, as dimensionality does not explicitly enter the preceding analysis. This is because, in the weak-field case, only the $\Gamma$-point modes are stable; therefore, physical quantities such as ionic conductivity and mean fields are independent of summations over the entire Brillouin zone. This behavior contrasts sharply with the mean-field analysis of electronic systems, where dimensionality plays a crucial role due to the necessity of integrating over all $\bm{k}$ points. As a brief verification, we analyze the two-dimensional honeycomb lattice in Appendix.~\ref{sec:extend_2D}. 

In conclusion, supported by the clear physical interpretation and unified theoretical framework developed in this work, we achieve a comprehensive understanding of the origins and material dependencies of the two types of superionic phase transitions, as summarized in Table~\ref{tab:comparison}.



\subsection{Extensions and Limitations}\label{sec:limitations_and_extentions}


It is worth noting several limitations of the preceding analysis. First, in the type-I superionic phase, the liquid-like motion of mobile ions signals the breakdown of the tight-binding approximation introduced in Sec.\,\ref{sec:approximations}; accordingly, our treatment does not extend into this regime. Nevertheless, even in the normal phase below the transition temperature, ionic dynamics exhibit strong anharmonicity, which can be more advantageous for thermoelectric performance than the fully superionic state. This is because long-range ionic diffusion in the superionic phase may enhance thermal transport via convective contributions and may also compromise structural stability~\cite{liu2025energy,ren2023extreme}. Hence, practical strategies for designing superionic thermoelectric materials may focus on stabilizing near-critical normal phases, where our model can still be applied. 

Furthermore, our discussion is based on a few-site unit cell, which restricts the analysis to high-symmetry materials with small unit cells and thus is not directly applicable to systems such as high-entropy materials~\cite{percolation_model} or garnet-type compounds~\cite{sigma_YB,nonmonotonic,sigma_G}. Additionally, the detailed dynamical response of the host lattice is not explicitly accounted for, such as the paddle-wheel mechanism~\cite{zhang2022exploiting}, which may generate more complex transition behavior when coupled to mobile ions.


\section{Conclusions}\label{sec:conclusion}

We have established a comprehensive theoretical framework for general solid-state ionic conductors. Building on stochastic and tight-binding approximations, we develop a statistical description of mobile-ion transport that is firmly grounded in microscopic lattice dynamics. By combining the adiabatic and single-particle approximations, our lattice model successfully captures the characteristic transport behavior observed in conventional solid-state ionic conductors. Using a self-consistent mean-field approach, we further incorporate the nonadiabatic coupling between mobile ions and the host lattice, as well as Coulomb interactions among the ions. These ingredients naturally give rise to type-I and type-II superionic phase transitions, respectively. Our model reproduces a broad range of experimental observations with high fidelity. Benefiting from a unified theoretical treatment of the two transition types, we also provide a comprehensive comparison and clarification of their distinct microscopic mechanisms. Overall, our results pave the way for advanced modeling and a deeper understanding of solid-state superionic conductors.

Moreover, we hope this work encourages greater research attention from the condensed-matter community to the rich collective phenomena in solid-state ionic systems. Although ionic transport typically exhibits classical behavior, this does not diminish the fundamental interest of these processes. As discussed in Sec.\,\ref{sec:statistical_description}, the statistical description of ionic transport bears an essential similarity to the statistical interpretation of quantum mechanics. The stochastic nature of mobile ions stems from incomplete knowledge of a macroscopic thermalized system, whereas the probabilistic behavior of quantum particles arises from the intrinsic wave nature of matter. This parallel echoes the well-known connection between quantum field theory and statistical field theory~\cite{weinberg1995quantum}, which are linked through the Wick rotation. Consequently, many powerful techniques originally developed for many-body or strongly correlated electronic systems, such as the self-consistent mean-field approach employed here, hold significant promise for advancing the study of ionic conductors and uncovering new emergent behavior. 


\acknowledgements

The authors acknowledge insightful discussions with Prof. Hua Wang (Zhejiang University), Prof. Qiyang Lu, Prof. Wei Zhu, Prof. Congjun Wu, Prof. Wenbin Li, Prof. Yizhou Zhu, Prof. Shi Liu, Dr Xiaoping Yao, and Mr Yunhao Liu (Westlake University), as well as with Prof. Bin Xu (Soochow University). J.H. thanks the China Scholarship Council for its support. B.M. acknowledges financial support from a UKRI Future Leaders Fellowship [MR/V023926/1].


\appendix

\section{Two-dimensional system}\label{sec:extend_2D}

\begin{figure}[htp]
    \centering
    \includegraphics[width=0.48\textwidth]{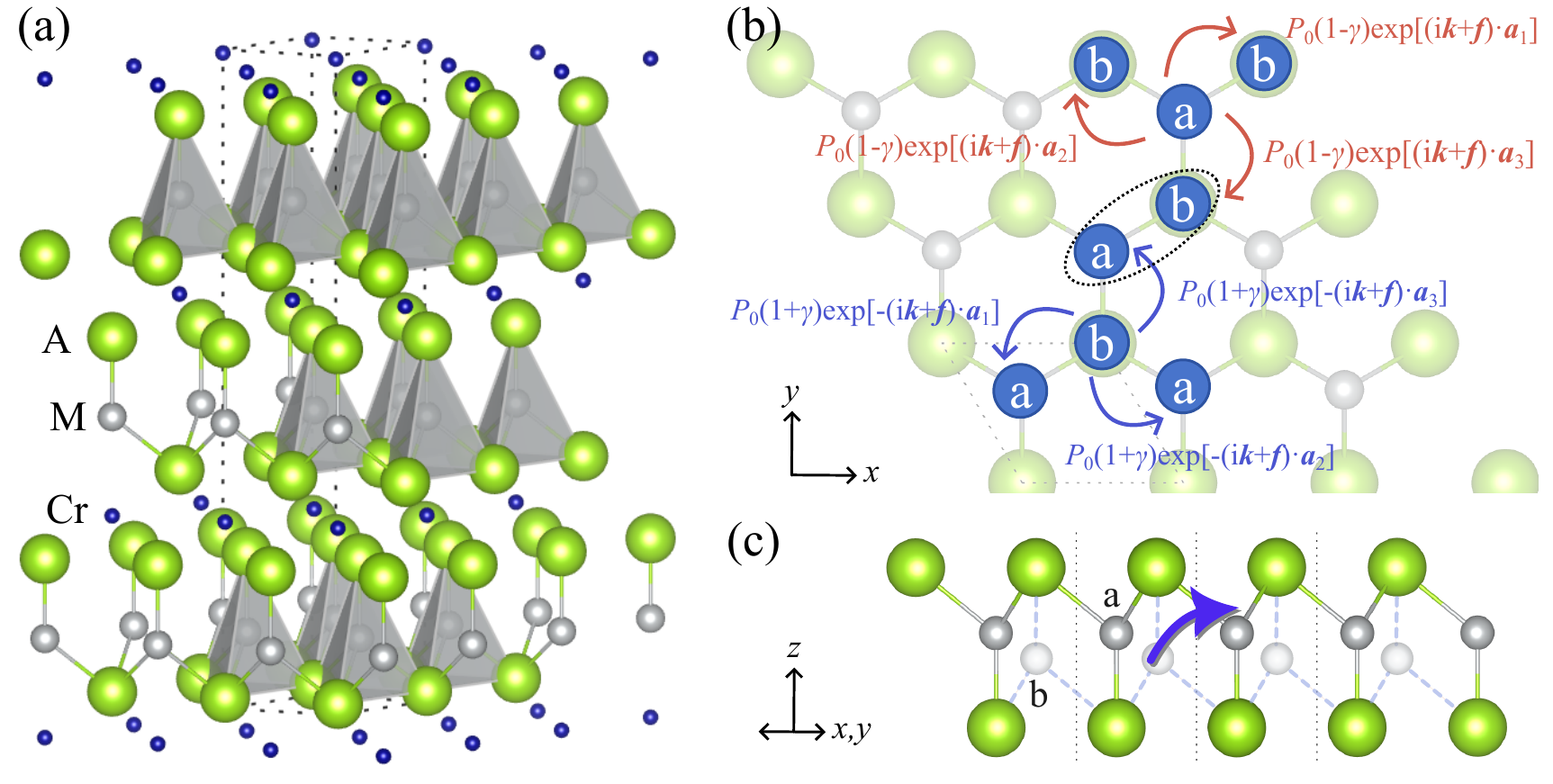}
    \caption{\label{fig:AgCrX2_structure}
    \textbf{Schematic of 2D honeycomb model.} (a) Structure of MCrX$_2$ (M=Cu,Ag; X=S,Se). (b,c) Schematic of nearest-neighboring ionic hopping. }
\end{figure}

Consider a specific two-dimensional mobile-ion system with a honeycomb lattice, which can represent the superionic material MCrX$_2$ (M=Cu,Ag; X=S,Se). The structure and hopping are presented in Fig.\,\ref{fig:AgCrX2_structure}. Similar to the 1D case discussed in Eq.\,\eqref{eq:Pk_1D_Efield}, the hopping matrix reads
\begin{equation}
    \begin{aligned}
        \mathbb{P}_{\bm{k}} &= 
        \begin{bmatrix}
            1-P_0(1 - \gamma) & P_0(1+\gamma)\Omega(\bm{k},\bm{f}) \\
            P_0(1-\gamma)\Omega(-\bm{k},-\bm{f}) & 1-P_0(1+\gamma) 
        \end{bmatrix}. 
    \end{aligned}
\end{equation}
Here $\bm{f}=Z\bm{\mathcal{E}}/k_{\rm B}T$ is the electric force weighted by the thermal energy. The structural factor is
\begin{equation}
    \begin{aligned}
        \Omega(\bm{k},\bm{f}) &= \frac{1}{3}\sum_{j=1}^3 e^{i(\bm{k}-i\bm{f})\cdot\bm{a}_j}, 
    \end{aligned}
\end{equation}
with inter-site vectors $\bm{a}_1=a(1/2, 1/2\sqrt{3}), \bm{a}_2=a(-1/2,1/2\sqrt{3}), \bm{a}_3=a(0, -1/\sqrt{3})$. The eigenvalues for the `$+,-$' modes are solved as
\begin{equation}
     p_{\bm{k},\pm} =  1 - P_0 \pm P_0\sqrt{\gamma^2 + (1-\gamma^2)\Omega(\bm{k},\bm{f})\Omega(-\bm{k},-\bm{f})}.
\end{equation}
In the weak-field limit, we still have a stable solution near the $\Gamma$ point, where the structural factor is approximated as: 
\begin{eqnarray}
    \begin{aligned}
        &\Omega(\mathbf{k},\mathbf{f})\,\Omega(-\mathbf{k},-\mathbf{f}) = 1 + i\frac{a^2}{3}( k_xf_x  + k_yf_y). 
    \end{aligned}
\end{eqnarray}
The eigenvalue of the stable `$+$' branch becomes:
\begin{equation}
    p_{\bm{k},+} =  1 + iP_0(1-\gamma^2)\frac{a^2}{6}( k_xf_x  + k_yf_y), 
\end{equation}
the frequency is: 
\begin{equation}
     \omega_{\bm{k},+} = \frac{ZP_0(1-\gamma^2)a^2}{6\tau k_{\rm B}T}( k_x\mathcal{E}_x  + k_y\mathcal{E}_y), 
\end{equation}
and the group velocity is:
\begin{equation}
    \begin{aligned}
        v^x_{0, +} &= \frac{\partial \omega_{\bm{k},+}}{\partial k_x} = \frac{ZP_0(1-\gamma^2)a^2}{6\tau k_{\rm B}T}\mathcal{E}_x,
        \\
        v^y_{0, +} &= \frac{\partial \omega_{\bm{k},+}}{\partial k_y} =\frac{ZP_0(1-\gamma^2)a^2}{6\tau k_{\rm B}T}\mathcal{E}_y.
    \end{aligned} 
\end{equation}

Therefore we get the conductivity tensor as: 
\begin{equation}
    \begin{aligned}
        \sigma_{xx} &= \frac{Zn_0}{V}\frac{\partial v^x_{0, +}}{\partial \mathcal{E}_x} = \left(\frac{J_0Za}{6k_{\rm B}T}\right)P_0(1-\gamma^2) = \sigma_{yy},
        \\
        \sigma_{yx} &= \sigma_{xy} = 0, 
    \end{aligned} 
\end{equation}
with a current density constant $J_0=\frac{Zn_0a}{V\tau}$. A comparison with Sec.\,\ref{sec:couple_E_1D} shows that, as stated in Sec.\,\ref{sec:relation_meanfields_1D}, the superionic transport properties in the weak-field limit are not sensitive to dimensionality.

\bibliography{references}

\end{document}